\RequirePackage{lineno}
\documentclass [preprint, nofootinbib] {revtex4} 
\usepackage{placeins}
\usepackage{booktabs}
\usepackage{multirow}
\usepackage{scrextend}
\usepackage{makecell}
\usepackage{graphicx}
\usepackage{amssymb}
\usepackage{amsmath}
\usepackage{color}
\usepackage{float}
\usepackage{afterpage}
\usepackage{xcolor}
\colorlet{darkred}{red!80!black}
\colorlet{darkgreen}{green!50!black}
\usepackage[normalem]{ulem}

\def\lp {\left( }
\def\rp {\right) }
\def\lb {\left[ }
\def\rb {\right] }

\def\bea{\begin{eqnarray}}
\def\eea{\end{eqnarray}}
\def\nn {\nonumber}

\def\d{\delta}

\def\l{\lambda}

\def\p {\pi}
\def\r{\rho}

\def\th {\theta}

\newcommand{\bppp}{B^+ \to  \pi^-\pi^+\pi^+}

\begin{document}

\title{ Suppressed $B \to PV $ CP asymmetry: CPT constraint}
\author{J.H.~Alvarenga Nogueira$^b$, I.~Bediaga$^a$, T.~Frederico$^b$, P.C.~Magalh\~aes$^a$, J.~Molina Rodriguez$^a$}
\email[]{pmagalhaes@cbpf.br}
\affiliation{$^a$Centro Brasileiro de Pesquisas F\'isicas, 22290-180, Rio de Janeiro, RJ, Brazil \\
$^b$ Instituto Tecnol\'ogico de Aeron\'autica, DCTA, 12228-900, S\~ao Jos\'e dos Campos, SP, Brazil}
\date{\today }

\begin{abstract}
Charge Parity asymmetry in charmless B meson decays is a key issue to be understood. 
Many theoretical calculations have been performed using short distance  factorization approaches which, in general,  do not take into account the  CPT invariance 
constraint. For each channel with CP violation there is an equal amount of CP asymmetry in another channel or other channels, with an opposite sign.  
This happens if these channels are coupled through final state interactions (FSI). In the specific process $B\to PV$, involving one pseudo-scalar 
and one vector particle in the final state, we argue that the CP asymmetry, inherent from a short distance mechanism, could be suppressed due to the CPT 
constraint. In this case, we propose a sensitive and practical experimental method to identify even a  small CP asymmetry, which provides
 the values for $A_{CP}$ without the need 
for an amplitude analysis. This method, if applied directly to data, will enable  to extract the CP asymmetry information in a model independent way and
check to which  extend  the suggested suppression due to the CPT constraint is verified.

 \end{abstract}

\pacs{...}

\maketitle

\section{Introduction}
\label{sec:intro}

Direct charge parity (CP) asymmetry in charmless heavy-meson decays is a long and intriguing puzzle that has been widely studied since the seminal BSS paper~\cite{BSS} 
in the late 70's. The idea, which became the BSS mechanism, is that the CP asymmetry comes from the interference between tree and penguin quark level 
diagrams owning  different weak and strong phases. 
 Nowadays, the most common concept used to compute branching fractions and CP asymmetries of charmless B meson decays is the factorization of the decay amplitude, which  was started with the Naive factorization approach (NF)~\cite{naiveFA}. The breakthrough came in the 80's by the seminal work of Lepage and Brodsky~\cite{brodsky}, which 
was the base to the main frameworks developed to study exclusive heavy meson decays.   For the B meson,  we recall that those main approaches are: 
QCD factorization (QCDF)~\cite{QCDF}, perturbative QCD (pQCD)~\cite{pQCD} and soft-collinear effective theory (SCET)~\cite{SCET}. 
All of them are based on the factorization of the hadronic matrix elements, changing technically how this is done, i.e.,  the ordering of the scales involved, treatment of 
dynamical degrees of freedom and  power counting. 
One example of the factorization approach with dressed hadronic form factors is given in~\cite{QCDFFSI}. 
There is also an alternative framework to factorization which considers the non-perturbative nature of the process based on the flavour $SU(3)$ symmetry~\cite{SU3F}.

In general these approaches considered only short distance amplitudes, without taking into account the constraint that CPT invariance imposes on the CP asymmetry at the 
hadronic level.    
For a long time, the common belief was that  heavy meson B decays would produce many hadronic channels with a homogeneous momenta distribution over the phase space,  
from the allowed rescattering process to several hadronic channels.
These possible FSI, involving two, three or more hadrons, induced the idea that CPT constraints would not be of practical use in charmless B decays~\cite{Bigi}.
 However, the high sample of experimental data, first with BaBar and Belle and then with LHCb, shows that  these multi-mesons rescattering 
 processes  are not dominant in charmless multi body B decays~\cite{BOT,ABAOT}.

Indeed, inspecting the LHCb data distribution of these decays~\cite{LHCb2014}, one can see  that the events are placed basically around the Dalitz plot boards, 
dominated by low mass resonances. The fact that charmless three-body B decays   populate regions close to  the edge of the Dalitz plot, reinforces the idea 
that those processes are dominated by two-body interactions.  In a recent theoretical paper~\cite{ManeG}  it  was shown
that these  data distribution of charmless three-body B decays is a consequence of the  relatively small  B meson mass.
 Following this paper, a homogeneous distribution in the Dalitz plot should be possible only for  three-body decays
 of particles with masses that are more than five times larger than the B meson.

In the  large compilation of B charmless processes presented in Ref.~\cite{chengchua2015}, one finds a quite reasonable agreement between  theories based 
on factorization techniques and  experimental results available for several  branching fractions of  $B \to PP$ and  $B \to PV$ decays. 
However, the same agreement is not seen for the values of CP asymmetries  in the same channels~\cite{chengchua2015}. 
In fact, there are many discrepancies not only among the different models but also comparing them to the experimental CP asymmetry measurements. 
 This situation is worse for $B \to PV$ decays. Nonetheless, it is important to note that there are some issues in these  channels for both theoretical 
 and experimental descriptions. In the latter, the processes $B \to PV$  are in fact three-body decays and consequently  the  observables  
 are determined  inside the complexity of a three-body phase-space.

In this paper we present  a general discussion of the theoretical issues in the $B \to PV$ decays, arguing that 
regardless of the model chosen, the CPT constraint can play a crucial role in suppressing the CP asymmetry in these channels.  
We also propose  a simple model to extract the CP asymmetry from these channels, avoiding the difficulty presented in Dalitz amplitude analyses, 
normally  used to extract the values of $A_{CP}$ in these  decays.

\section{ CPT and CP asymmetry suppression in  $B\to P V$ decays }
\label{sec:cptconst}
In literature, the theoretical studies for  $B \to PV$ decays are limited to  low mass SU(3) vector particles: $\rho(770)$, $K^*(890)$ and $\phi(1020)$. 
Both  $\rho(770)$  and $K^*(892)$  decays  are restricted to one single two-body channel and
the branching fraction for their decays are 100\% to $\pi\pi$ and to $K\pi$, respectively \cite{pdg}.
The case of $\phi(1020)$ is slightly different, it decays  mainly into $K \bar{K}$ but  it has also a decay involving three pions through the 
$\p\rho(770)$ channel with a branching fraction of 15\% \cite{pdg}.
 
The isobar model, as well as K-matrix and other amplitude analyses, widely used by experimentalists for three-body heavy meson decays,
considers the bachelor particle as a simple spectator of the process. In this quasi-two-body  process or  (2+1) approximation, 
resonances  produced in heavy meson decays do not interact with the third particle.  Within this scenario, the  CPT constraint 
 in $B\to P V$ processes, where $V=$ $\rho(770)$ or $K^*(890)$, suggests that there is no room to observe CP asymmetry in these channels.
There is also the $\phi(1020)$ resonance, which could show CP asymmetry, but that does not seem possible due to the low contribution of tree diagrams 
to decays involving this vector particle.  It is worth mentioning that the absence of final state interactions is a hadronic constraint and therefore, the 
impossibility  to observe CP asymmetry in those processes is independent from the relative short distance contribution from tree and penguin diagrams.

 There are at least three  
other possibilities for $B \to PV$ decays resulting in inelastic rescattering that can produce CP violation: \\
(i) rescattering from the pseudoscalar-vector like $P V \to P' X'$, where X is a new particle or particles. 
\\
(ii) The  $P V $ final state was not produced promptly but is a result of a rescattered process coming from another two-body decay channel, with the strong transition matrix related with the one in (i) 
by detailed balance, or time reversal invariance. \\
(iii) A three-body rescattering including the bachelor particle. 

The CPT constraint demands that the total 
CP asymmetry  distributed in different channels or phase-space regions coupled with the strong interaction add to zero.
These processes (i)-(iii) are estimated to provide small
contributions  to the  CP asymmetry distribution for the $B$ meson decay in different channels coupled by the strong 
interaction. The flux of CP asymmetry between channels/phase space regions is
proportional to the  S-matrix  expectation value between the different states.

The probability calculation of a light-meson  processes (i) or a transition amplitude for
 (ii) $P V\to P'X'$ or $P' V'\to PX$ should, in principle, start  with the QCD theory, which is much beyond the scope of this work. 
 Nevertheless, a rough estimation of the transition amplitude 
for a  light-meson pair to scatter into a different pair of mesons at the $B$ mass is enough
to exclude the importance of the coupling between two-body channels at such energies, as long as unitarity is respected. 
The off-diagonal S-matrix elements should have a magnitude lower than one due to unitarity.
Its modulus square can be interpreted as the  probability for the transition between the initial and final 
channels. Then, our task is to estimate the Lorentz invariant matrix element for the transition matrix associated
to the inelastic scattering process. 

The naive picture for two-body inelastic collision of two initial mesons
is the annihilation of these two hadronic 
states to a  quark-antiquark pair that propagates and recombine  producing the light-meson pair
in the final state. The contribution from the  intermediate state propagation of the quark pair 
in the Mandelstam variable $s$  is damped roughly by a factor $s^{-1}$. In addition, the  breakup of a 
meson in a quark-antiquark pair has to bring a damping factor of $s^{-1}$, because in this case there is an imbalance 
in the relative momentum of the quarks by  $\sim \sqrt{s}$. If in one of the mesons at the initial or final state 
the relative momentum of the pair is small then the other one should compensate with a large relative momentum,
 which is damped by the wave function of the meson. The asymptotic behaviour 
of the lowest Fock-state  component of the light-cone wave function in $S-$wave decreases with the transverse 
momentum as $k^{-2}_\perp$ \cite{JiPRL03}, which we naively associate with the invariant $s^{-1}$.
Therefore, the Lorentz invariant matrix element of the transition operator should carry a damping factor of at 
least $\sqrt{s-s_{th}}/s^{3.5}$, where the threshold behaviour was included without changing the asymptotic behaviour 
in $s$. 

 According to the above discussion, the damping factor in the off-diagonal S-matrix element 
corresponding to  $P V\to P'X'$ inelastic collision, is estimated as
\begin{equation}\label{sinel}
S_{P V\to P'X'}(s)\sim \mathcal{N}\sqrt{s/s_{th}-1}\,/{(s/s_{th})^{\alpha}}, 
\end{equation}
with $\alpha=3.5$.  The S-matrix element  cannot be larger than 1 due the unitarity constraint. 
By choosing  the 
normalization factor as $\mathcal{N}=\Lambda^6=(1.24)^6$ in Eq.(\ref{sinel}),  the  maximum  value reaches $\sim 0.87$,  
when $\sqrt{s}\,=\,1.08\, \sqrt{s_{th}}$. An example for the application of the formula (\ref{sinel}) is  
the $s-$wave isospin zero $\pi\pi\to KK$ cross section, which drops fast and is relevant below $\sqrt{s}\sim$ 1.6 GeV~\cite{cern-munich}. This naive formula is consistent with  one of the parametrizations presented in~\cite{PelPRD05} for the 
inelasticity parameter $\eta(s)=\sqrt{1-|S_{\pi\pi\to KK}(s)|^2}$.
 Using this simple formula with $\sqrt{s}_{th}=2$ GeV,  it results in $S_{P V\to P'X'}(m_B)\sim 0.014$, 
which provides a large suppression of the off-shell amplitude  at the $B$ meson mass.

In process (iii), namely a three-body rescattering process,  one of the  
pseudo-scalar meson scatters with the bachelor particle and, therefore, the CP asymmetry, can flow 
from one region of the three-body phase-space to another. At the amplitude level this one-loop process introduces new complex structures that 
could affect different partial waves.  
 This effect was studied in the context of the charged $D^+\to K^-\pi^+\pi^+$ decay \cite{patriciaPRD, karinJHEP, patWeak}, and the results 
indicate that it was crucial to explain the observed experimental  S-wave  phase-shift. 
It was also shown that it is suppressed by a factor of about 20\% with respect to the driving term. 
If we roughly estimate that it has
a dependence on $s^{-1}$, then from the D decay to B decay, one has a suppression factor $\sim (1.89/5)^2$ giving
$\sim 5\%$ for the three-body rescattering contribution with respect to the driving partonic amplitude. 
Recently, we performed a qualitative study~\cite{patIg} about the importance of FSI rescattering to $\bppp$ decay  suggesting 
that the presence of hadronic loops shifts the P-wave phase near the threshold to below zero, and modifies the position of the 
$\r$-meson peak as well as its width, in the Dalitz plot. 
Basically, the obtained contribution for the effect was a small  percentage, but a more refined analysis which also considers  the S-wave rescattering is 
being developed and should be released soon \cite{pat2016}.

 However, the amplitude analyses used to extract CP- asymmetry information from the data are model dependent and based on a (2+1) approximation. 
Thus, there is still room for improvement as will be discussed in the following section. 

\section{ Model independent method to extract $A_{CP}$ in  $B\to P V$ decays}
\label{sec:model}

To avoid the dependence of the isobar model when extracting  parameters of CP asymmetry in  $B\to P V$  processes, we propose a simple  model independent experimental
 procedure to extract the $A_{CP}$ from the data. The method explores the angular distribution of a  vector resonance in the 
 Dalitz plot and the property that the low mass vector meson is in general  close to a scalar one, sharing the same region of the phase space.
Thus, one can take a slice from the central mass of a light vector resonance that includes the interference with  only one low mass scalar resonance along the other Dalitz variable. These scalar resonances  can be the $\sigma$, $\kappa$, $f_0(980)$  or even a non-resonant contribution.

To illustrate the method we begin with a simple situation  where $B^\pm \to h^\pm (V \to h^+ h^-)$ decay (h=$\pi$ or K)  receives contributions only from one vector resonance (V) and a constant nonresonant (NR) amplitude. 
 Generically, one could represent the total amplitudes  for $B^{+} $ and $B^{-}$ charge conjugate decays  as:
\bea
{\cal M}_+ &=& a^{V}_+ e^{i \delta^{V}_+}F^{BW}_{V} \cos \theta({s_{\perp}},s_\parallel) + 
a^{nr}_+ e^{i \delta^{nr}_+}F^{NR}, 
\label{RHOAMPNR1}\\
{\cal M}_- &=& a^{V}_- e^{i \delta^{V}_-}F^{ BW}_{V} \cos \theta({s_{\perp}},s_\parallel) + 
a^{nr}_- e^{i \delta^{nr}_-}F^{ NR}, 
\label{RHOAMPNR2}
\eea
where the $F^{ NR}$ is a real and scalar non-resonant amplitude, and $\delta _{\pm}$ contains both the fixed weak and  strong phases.  The vector resonance V is described by a Breit-Wigner (BW) function, $F^{BW}_{R}$, that depends on $s_\parallel = (p_{h_+} + p_{h_-})^2 $, one of the invariant variables at Dalitz plot,
\begin{eqnarray}
F^{ BW}_{V} (s_{\parallel}) = \frac{1}{m^2_{V} - s_{\parallel} - i m_{V}\Gamma_{V}(s_{\parallel})} ~,
\label{BW}
\end{eqnarray}
and  $\Gamma_{V}(s_{\parallel})$ is the energy dependent relativistic width.
The vector amplitude has   an additional strong phase, inherent to the BW form, and a spin 1 factor, proportional to $\cos \theta(s_{\perp},s_\parallel)$. 
The angle $ \theta $ is defined as the helicity angle between the bachelor and  center of mass of the two particles produced by the resonance. 
For a vector resonance in the $s_\parallel$ channel, the cosine of helicity angle is given by~\cite{Kajantie}
\bea
\cos \theta(s_\perp, s_{\parallel}) &=& \frac{(M^2_B  - s_\parallel - M^2_{h_b})( s_\parallel + M^2_{h_+} - M_{h_-}^2 ) + 2\,s_\parallel (M_{h_b}^2 +  M^2_{h_+} - s_\perp)}{\sqrt{\l(M^2_B, s_\parallel, M^2_{h_b})}\,\sqrt{ \l( s_\parallel, M^2_{h_+}, M^2_{h_-})}}~,
\label{cos}
\eea 
where $\l(x, y, z)$ is the Kallen function and    
{\small $ \sqrt{\l(x, y, z)}= \lb x - \lp \sqrt{y} + \sqrt{z}\rp ^2\rb \lb x-\lp \sqrt{y} - \sqrt{z}\rp ^2\rb \,$}.
We investigate the behaviour of $\cos \th(s_\perp,s_{\parallel})$   for values of $s_\parallel \approx m_V^2$, where $V =\r(770)$, $K^*(892)$ and $\phi(1020)$. We  found that the values of $\cos \th$ remain stable as a function of $s_\perp$  around the center value of $s_\parallel=m^2_V$, i.e.,  within a  region of about the width of the resonance. Therefore,  the helicity angle can be assumed to be a  function of only $s_\perp$: $\cos \th ( s_\perp,m^2_V\pm \d_m) \approx \cos\th(s_\perp,m^2_V) $.

The CP asymmetry is obtained from the ratio of  subtracting/adding the square modulus of the $B^+$ and $B^-$  amplitudes: 
\bea
{|\cal M}_+|^2 \mp |{\cal M}_- |^2 &=&[ (a^{V}_+)^2 \mp (a^{V}_-)^2] |F^{ BW}_{V}|^2
\cos^2 \theta({s_{\perp}},s_{\parallel})+ [ (a^{nr}_+)^2 \mp (a^{nr}_-)^2] |F^{NR}|^2 
\nn\\[2mm]&&+ 2 \cos\theta(s_{\perp},s_{\parallel}) |F^{ BW}_{V}|^2|F^{ NR}|^2 \: \times \nonumber \\ 
&& \qquad \{(m_{V}^2 -s_{\parallel})) \:  
[a^{V}_+a^{nr}_+ (\cos(\delta^{V}_+ - \delta^{nr}_+) \mp a^{V}_-a^{nr}_- \cos(\delta^{V}_- - \delta^{nr}_-)] \nonumber \\[2mm] 
&& \qquad \quad -m_{V}\Gamma_{V} \: \: [a^{V}_+a^{nr}_+ (\sin(\delta^{V}_+ - \delta^{nr}_+) \mp a^{V}_-a^{nr}_-\sin(\delta^{V}_- - \delta^{nr}_-)]\}. 
\label{DeltaMNR}
\eea
 A  formula similar to Eq.~(\ref{DeltaMNR}), but integrated in $s_\perp$, was used in~\cite{ABAOT} to explore different types of CP asymmetries in 
charmeless three-body B decays. The formula was applied in the fitting of the integrated 
experimental distributions released by the LHCb experiment~\cite{LHCb2014}.  Here, instead, we propose  to look for the distribution on the $s_{\perp}$
 variable around the resonance mass, i.e., $s_\parallel \approx m^2_V$.

The usefulness of this procedure is that it enables us to identify the signature of $\cos \theta(s_\perp,m^2_V)$  when looking at the amplitude distribution in 
$s_\perp$. By doing that, one can relate the cosine  signatures to a specific type of CP asymmetry  source. Inspecting Eq.~(\ref{DeltaMNR}), one notes that the 
 first two terms  are associated to the direct CP asymmetry, created from  BSS mechanism. The former is related to the vector resonance and is  proportional to
  $\cos^2 \theta(s_\perp,m^2_V)$, whereas the second  is constant and associated to the scalar NR amplitude. The last two terms in 
Eq.~(\ref{DeltaMNR}) are proportional to  $\cos \theta(s_\perp,m^2_V)$ and related to the interference between the NR amplitude  and the vector resonance.
 This type of  CP asymmetry has two contributions: one associated to the real part of the vector amplitude and the other to the imaginary. 

In other words,  the  coefficients $(a^{V}_\mp)^2$  tracks the $\cos^2 \theta(s_\perp,m^2_V)$, which is related to a CP 
asymmetry from the BSS mechanism on the vector meson decay amplitude.
 The coefficients $a^{V}_\mp$ tracks the linear representation of the $\cos \theta(s_\perp,m^2_V)$ and is related to a CP 
 asymmetry produced from the FSI interference. 
 Finally, the coefficient $(a^{NR}_\mp)^2$ represents the possibility of CP asymmetry  produced by BSS mechanism in the 
  NR amplitude, related to the second term in Eq.~(\ref{DeltaMNR}). All these coefficients can be obtained directly from data 
  by fitting the square amplitudes with a quadratic function
   on $\cos \theta(s_\perp,m^2_V)$.
  
  The constraint of CPT applied to the CP asymmetry in the hypotheses of
   no three-body rescattering contributions, as discussed in the previous section,  imply that the integral over the phase space of the asymmetry, computed with Eq. (\ref{DeltaMNR}),  should vanish as discussed in detail in ~\cite{ABAOT}. 
   Concerning the linear term in $\cos\theta$, independently of the context it will  vanish  after integration over 
   the phase-space once it is an odd-function.  Therefore, CPT implies that the integration over the phase-space of the first two terms of 
   Eq. (\ref{DeltaMNR}) must be zero. The most probable solution is that the coefficients from $B^+$ and $B^-$ are the same. However, they 
   could also be different and compensate upon integration over the phase-space.
   
An important product of this method is the possibility of extracting the values of  $A_{CP}$  in a direct measurement of the CP asymmetry inherent to the BSS mechanism, 
within a model independent approach.  From the  quadratic coefficients  of the fit one directly obtains   the $A_{CP}$  without needing a model for the amplitude:  
\bea
A^V_{CP} =  \frac{(a^{V}_-)^2  - (a^{V}_+)^2 }{(a^{V}_-)^2 + (a^{V}_+)^2}.
\label{Acp_fit}
\eea

The inclusion of  one low mass scalar resonance like $\sigma$, $\kappa$ or $f_0(980)$, does not change the main features of this method.
 In this case, the charmless three-body B decay amplitudes are given by replacing the NR amplitude for a Breit-Wigner denoted as $F^{\rm BW}_{S}$. 
The scalar resonance has an inherent strong phase that will interfere with the vector amplitude. 
 The CP asymmetry considering the scalar resonance results in
\begin{eqnarray}
|\Delta{\cal M}|^2 &=&{|\cal M}_+|^2 - |{\cal M}_- |^2  \nonumber\\
                   &=&[ (a^{V}_+)^2 - (a^{V}_-)^2] |F^{\rm BW}_{V}|^2\cos^2 \theta+ [ (a^{S}_+)^2 - (a^{S}_-)^2] |F^{\rm BW}_{S}|^2 +\,2\,\cos\theta |F^{\rm BW}_{V}|^2|F^{\rm BW}_{S}|^2 \times \nonumber \\
&& \{[ (m_{V}^2 -s) (m_S^2-s) - m_{V}\Gamma_{V} m_{S}\Gamma_{S}][a^{V}_+a^{S}_+ \cos(\delta^{V}_+ - \delta^{S}_+) - a^{V}_-a^{S}_- \cos(\delta^{V}_- - \delta^{S}_-)]\nonumber \\
&& -[m_{V}\Gamma_{V}(m_S^2-s) - m_{S}\Gamma_{S}(m_{V}^2 -s))
[a^{V}_+a^{S}_+ \sin(\delta^{V}_+ - \delta^{S}_+) - a^{V}_-a^{S}_- \sin(\delta^{V}_- - \delta^{S}_-)]\}\,.\nonumber\\
\label{DeltaMfrho}
\end{eqnarray}
Comparing this  formula with Eq.~(\ref{DeltaMNR}), one notes that the coefficients related to BSS  mechanism  are the same and the interference term is also 
proportional to $\cos \theta(s_\perp,m^2_V)$. Therefore, the amplitudes can  be parametrized by the same quadratic function of $\cos \theta(s_\perp,m^2_V)$ as in 
the previous example.
%
 
The data scenario is more complex because there are also  resonances placed in a crossed channel, that are functions of  $s_{\perp}$, and can interfere in a non trivial form with  the distribution of $\cos \theta(s_\perp,m^2_V)$, producing other sources of CP asymmetry.  However these interferences are localized at  lower masses
(in general below 5 GeV$^2$). Taking into account the big  phase space accessible for charmless three-body B decays, even if we exclude this interference region, there is still a large region to perform the analysis  and extract  the  $A^V_{CP}$ measurements from the  fit parameters, with good resolution and limited errors.

 \FloatBarrier
\section{Viability of method  through fast Monte Carlo simulation }
\label{sec:toymc}

In this section, we  apply the previously presented method for the $B^\pm \to \p^\pm \p^+\p^-$ decay generated by using Toy Monte Carlo (MC) pseudo-experiments. 
The samples for these studies were simulated from  the results obtained in the BABAR experiment~\cite{babar}, by fitting the data  with an isobar model which  includes: the vector resonances $\r(770)$ and $\rho(1450)$, the scalar  $f_0(1370)$, the  tensor $f_2(1270)$ and a flat nonresonant contribution.  In their fitting results,   they had obtained a CP asymmetry in  all the  resonance  channels considered, including the NR contribution. Specifically for the channel  we are interested in, $B^\pm \to \rho(770) \p^\pm $,   they found an $A_{CP}^{\rho(770)}=18\% \,\,\pm \,7\%$.

In order to show the viability of the method employed to identify the CP asymmetry signatures,  we have considered two different scenarios for the Toy MC: one of them using  the BABAR inputs for magnitude and phase, and a second one where we only modify the magnitudes for $\r(770)$  in order to produce an $A_{CP}^{\rho(770)}=0$.  
For both situations we have generated 1000 samples by using the Laura++~\cite{laura} package,
 each sample with 20,000 events. By  applying  the method we aim, in the first case,  to reproduce the $A_{CP}^{\rho(770)}=18\%$ obtained by Babar, whereas, for the second case, we should  find $A_{CP}^{\rho(770)}=0$.

The procedure to calculate the $A_{CP}$ is straightforward: we choose a 50 MeV mass window around  the vector resonance $\rho(770)$, 
  integrating it  in the parallel Dalitz variable, and then fit the binned distributions of $B^+$ and $B^-$ histograms in the orthogonal 
  Dalitz variable with a quadratic polynomial function of $\cos \theta(s_\perp,m^2_V)$ 
  in the interval of  5 to 23.5 GeV$^2$. Following that,  we calculate the $A_{CP}$ by applying the quadratic coefficients previously 
  obtained from both fitted curves, and replacing its values into Eq.~(\ref{Acp_fit}). 
It is worth mentioning that the choice of the fitting interval excludes some interference effects. 
As we have argued before, these effects result from the  $\rho(770)$ inference with the crossed channel, which produces an extra CP asymmetry that does not come from the BSS mechanism.  Therefore, it is important to exclude these regions. 

In short, with this procedure we find  $A_{CP} = 0.177 \pm 0.03$  and  $A_{CP} =  0.0 \pm 0.03$ for the two cases under study. The results for the  fit with  the quadratic function are given in  Fig.~1 and the values for the quadratic coefficients are  summarized in Table~\ref{tab:vals}. We do not report linear and constant 
 fit parameters, once they do not add relevant information to the aims of our study.   

Besides the simplicity of the method  proposed,  the results for the Toy MC study are very satisfactory.  
The values for $A_{CP}$ obtained  from  the fitting coefficients  agree with  the correspondent  input values within the  
statistical errors. We also observe that although we restrict our study to a small  part of the three-body B phase space, 
the $A_{CP}$  errors of these samples are competitive with the ones extracted by an amplitude analysis with a similar sample. 
Thus, our main conclusion is that this method is reliable and easy to be applied to data.

\begin{figure}[H]
\begin{center}
{
\includegraphics[width=7.5cm,height=5.cm]{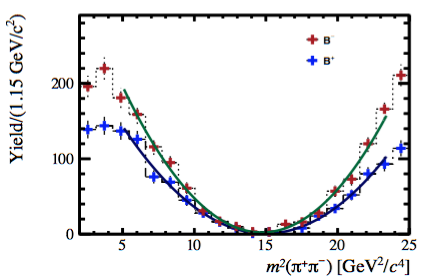}
\includegraphics[width=8.0cm,height=5.cm]{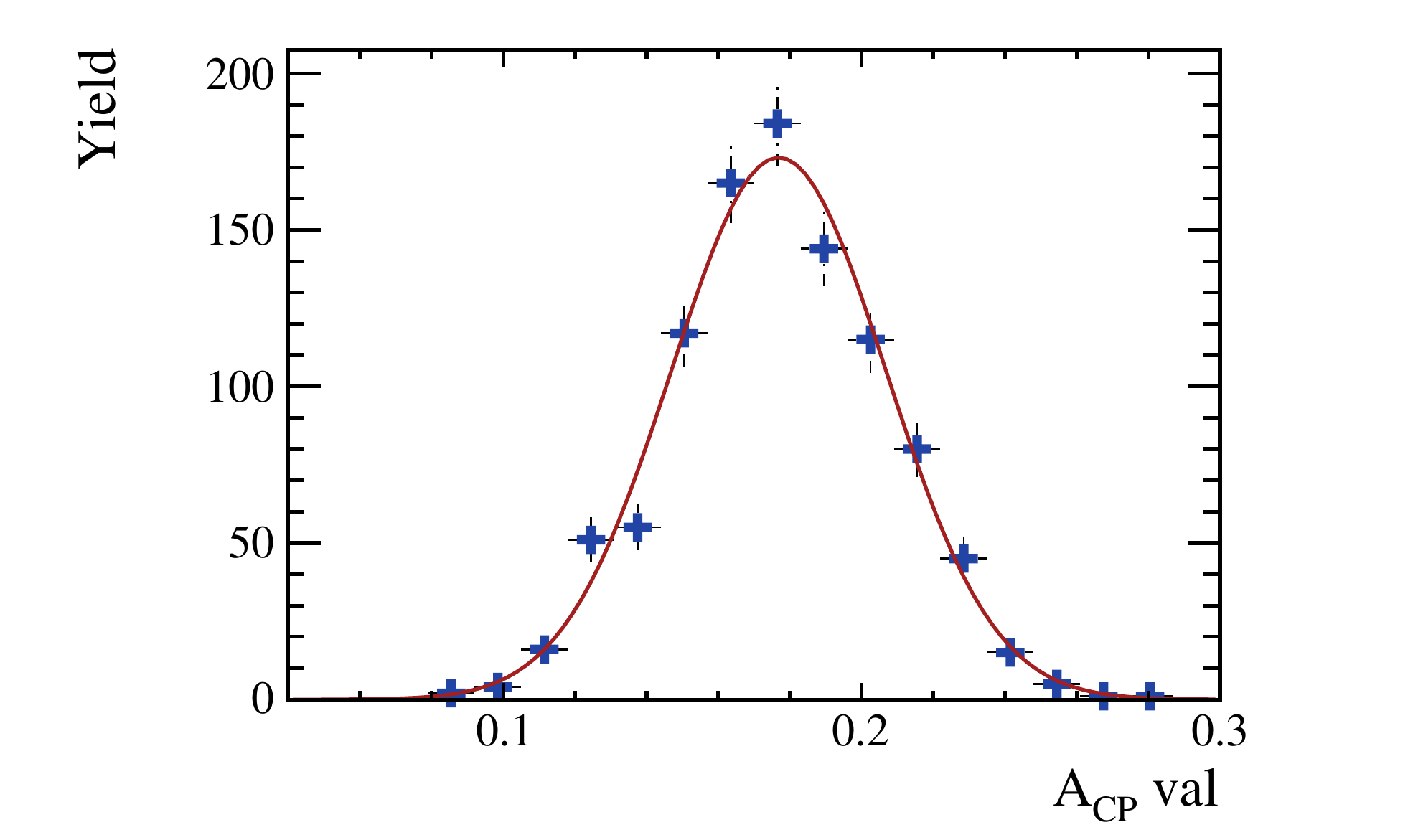}\\

\includegraphics[width=7.5cm,height=5.cm]{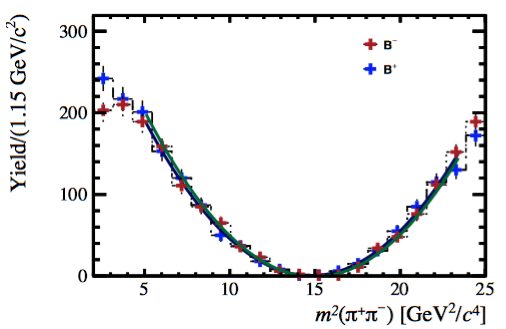}
\includegraphics[width=7.9cm,height=4.9cm]{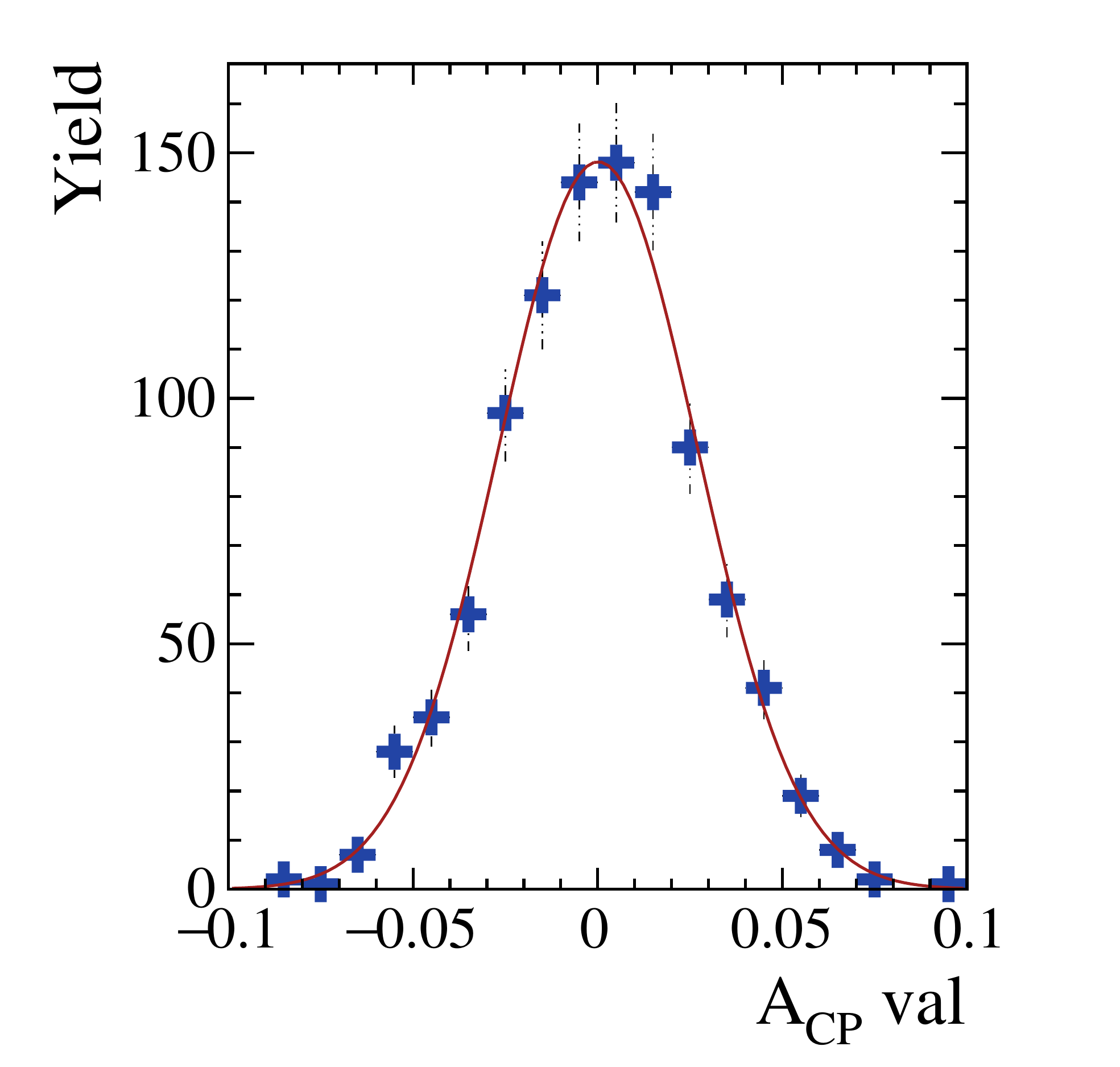}
{\small \caption{ Left:  fit to a random sample of binned distributions of $B^+$ and $B^-$ histograms generated with  $A_{CP}^{\rho(770)} = 0.18$ (top)
and  $A_{CP}^{\rho(770)} = 0.0$ (bottom). Right: the distributions of the main value for $A_{CP}$ obtained from the fits to the 1000 pseudo-experiments 
with 20000 events each, simulated with:  the BABAR results  (top), and   $A_{CP}^{_\rho(770)} = 0.0$ (bottom). 
}}
}
\end{center}
\label{fig:babar_1}
\end{figure}

\begin{table}[htpb!] 
\begin{center}
\begin{tabular}{c|ccc}
\multicolumn{1}{c}{}&  \multicolumn{ 2}{c}{$A_{CP}$} \\
 \specialrule{1pt}{0pt}{0pt}
\multicolumn{1}{c}{}& 0.18        & 0.0    \\ 
\toprule   \hline  
 $a^2_-$& $ 1.98\pm0.075$   &  $1.99\pm 0.077$  \\
  $a^2_+$ &$ 1.41\pm0.063$   &  $ 1.99\pm 0.076$ \\
    \hline
\bottomrule
\end{tabular}
\caption []{Central values for the quadratic coefficients resulting from the  fits of the 1000 samples generated with the given $A_{CP}$ with the quadratic function on $\cos \theta(s_\perp,m^2_V)$. The values for $A_{CP}$  can be extracted from the method by including the coefficients $a^2_-$ and  $a^2_+$ in Eq.~(\ref{Acp_fit}).}
\label{tab:vals}
\end{center}
\end {table}

\FloatBarrier
\section{Final remarks}
\label{sec:conclu}

Although short distance approaches in theoretical calculations indicate a substantial contribution to the direct CP asymmetry  in some charmless  $B\to PV$ decays, CPT constraints can suppress these possible contributions. Indeed, there are two experimental features involving these decays that reinforce those constraints.
 First, the hadronic three-body final state observables from charmless B decays have their events  placed on the edge of  the Dalitz plot. 
This  favours a (2+1) approximation with two-body resonances  plus a spectator particle. 
 Second, the elastic dominance in the $\rho(770)$ and  $K^*(890)$ resonances suppresses the coupling of these channels through FSI. Combining these two features  in the limit where the (2+1) approximation would be exact,  one can  reasonably consider that there is no possibility to have a direct CP violation in $B\to PV$ channels, since  there is no inelastic rescattering to satisfy the CPT constraint. 
However, there is still room for  inelastic rescattering contributions that could generate a  direct CP violation from the BSS mechanism, even though they are suppressed when compared with  short distance calculations.   

In order to perform $A_{CP}$ measurements for  several charmless $ B\to PV$ decays, we have proposed a model independent procedure to extract these experimental values. The method used some important charmless three-body  B decay properties, such as the large phase space, the relatively low mass of  $\rho(770)$, $K^*(890)$ and $\phi(1020)$  resonances, and  their  exclusive interference with scalar amplitudes along  all of the phase spaces. The Toy Monte Carlo simulation, based on experimental amplitude analysis results, proved the  utility of the proposed method, which succeeded in extracting the correct values of $A_{CP}$ with limited  errors. Moreover, the information one extracts from this method directly confronts several theoretical predictions for $A_{CP}$ \cite{chengchua2015}. 

It is worth noting that this  method has been successful in its purpose, which consists  in  measuring directly and model independently the $A_{CP}$  in $B\to PV$ processes, avoiding the complexity of amplitude analyses for three-body decays. The latter  is a tool to access  the resonance branching fractions  -  beyond the reach of our method -  providing  a complete description of three-body decays once it  deals with many other resonance structures and their interferences in a model dependent way.
 Besides this, both approaches can be complementary.
One can see that the method we presented here is very sensitive in detecting interferences in  the  crossed channel, that are functions of  $s_{\perp}$, as we can see in our toy Monte Carlo studies. This feature can be used to investigate the possibility of new high mass resonances in the  $s_{\perp}$ invariant mass, which can only be confirmed - including their specific properties -  through the use of the amplitude analysis technique.

Furthermore, it would be important to investigate the possibility of the rescattering of double charm B decays into light pseudoscalar mesons as proposed by Wolfestein\cite{wolfenstein}, in order to explain part of the possible dynamics of direct CP violation involving charmless three-body B decays.  

\section*{Acknowledgments} 

We wish to thank Fernando Ferreira Rodrigues for the fruitful discussions. We also thanks the support from Conselho Nacional de Desenvolvimento Cient\'ifico e 
Tecnol\'ogico (CNPq) of Brazil. The work of J.H.A.N. was supported by the grant $\#$2014/19094-8 from S\~ao Paulo Research Foundation (FAPESP).

\FloatBarrier

\end{document}